\documentclass{appolb}

\usepackage{epsfig}
\usepackage{amssymb}
\usepackage{amsmath}
\usepackage[ansinew]{inputenc}

\usepackage{soul,color}
\usepackage{graphicx}
\usepackage{dcolumn}
\usepackage{bm}

\usepackage{float}
\usepackage{mathrsfs}
\usepackage{tabularx}
\usepackage{midfloat}

\newcommand{\be}{\begin{equation}}
\newcommand{\ee}{\end{equation}}
\newcommand{\bea}{\begin{eqnarray}}
\newcommand{\eea}{\end{eqnarray}}
\newcommand{\ba}{\begin{array}}
\newcommand{\ea}{\end{array}}
\newcommand{\bi}{\begin{itemize}}
\newcommand{\ei}{\end{itemize}}
\newcommand{\lan}{\langle}
\newcommand{\ran}{\rangle}

\begin{document}
\title{Nucleon Resonances in Nuclear Matter and Finite Nuclei}
\author{Horst Lenske
\address{Institut f\"{u}r Theoretische Physik, Justus-Liebig-Universit\"{a}t Giessen, Giessen, Germany}
}
\date{\today}
\maketitle

\begin{abstract}
The theory of nuclear  excitations involving nucleon resonances is revisited and significantly extended to  asymmetric nuclear matter and higher P- and S-wave $N^*$ resonances. Excited states of are described as superpositions of particle--hole configurations including $NN^{'-1}$ and $N^*N^{-1}$ configurations. Configuration mixing is taken into account on the one--loop level by solving the generalized $N^*RPA$ Dyson equation. The underlying coupled channels formalism is derived and response functions is discussed. Applications of the approach are illustrated for charge--exchange modes of asymmetric nuclear matter and finite nuclei. The spectral gross structures of corresponding excitations in finite nuclei are investigated in local density approximation. Applications of the approach to resonance studies by high-energy heavy ion reactions are recapitulated.
\end{abstract}

\tableofcontents

\section{Introduction}\label{sec:Intro}

An exciting feature of nuclei is their large variety of dynamical modes, covering single particle degrees of freedom to quasi-elastic (QE) collective rotations and vibrations, passing over into the resonance (RE) region with excitations of the medium involving excited nucleons and entering the deep-inelastic sector. While single  particle and quasi-elastic modes probe merely the collective properties of a quantummechanical many-body system governed by the low-energy limit of strong interactions, in-medium resonance excitations penetrate into the sub-nuclear regions, offering unique opportunities to study the polarization of the nucleon in the dense environment of nuclear matter. In the interior of a heavy nucleus nucleons are densely packed with an average separation of $r_0\sim 1.14$~fm which is about the diameter of a nucleon. Nuclear many-body theory has developed a sophisticated framework of methods how to deal with these peculiarities, ranging from mean-field dynamics and collective models to self-consistent microscopic approaches allowing to describe quite precisely nuclear masses and ground state properties and the response functions observed by probing nuclei with various external probes.

Quite recently, peripheral heavy ion charge exchange reactions, probing hadronic charged current (CC) modes of nuclei. were performed at the FRS fragment separator at GSI - and will be continued at the Super-FRS at FAIR. Heavy ion bemas of 1~AGeV incident energy were used to study both types of CC response, $\tau_+$ modes populating $pn^{-1}$ configurations, and the complementary $\tau_-$ modes populating $np^{-1}$ configurations. In a heavy ion reaction a
$\tau_+$ process in the one nucleus is accompanied by a $\tau_-$ transition in the other nucleus, as required by charge and isospin conservation. In a nucleus A(Z,N) with Z protons and N neutrons, however, isospin symmetry is spontaneously broken.

\begin{figure}
\begin{center}
\includegraphics[width=6cm]{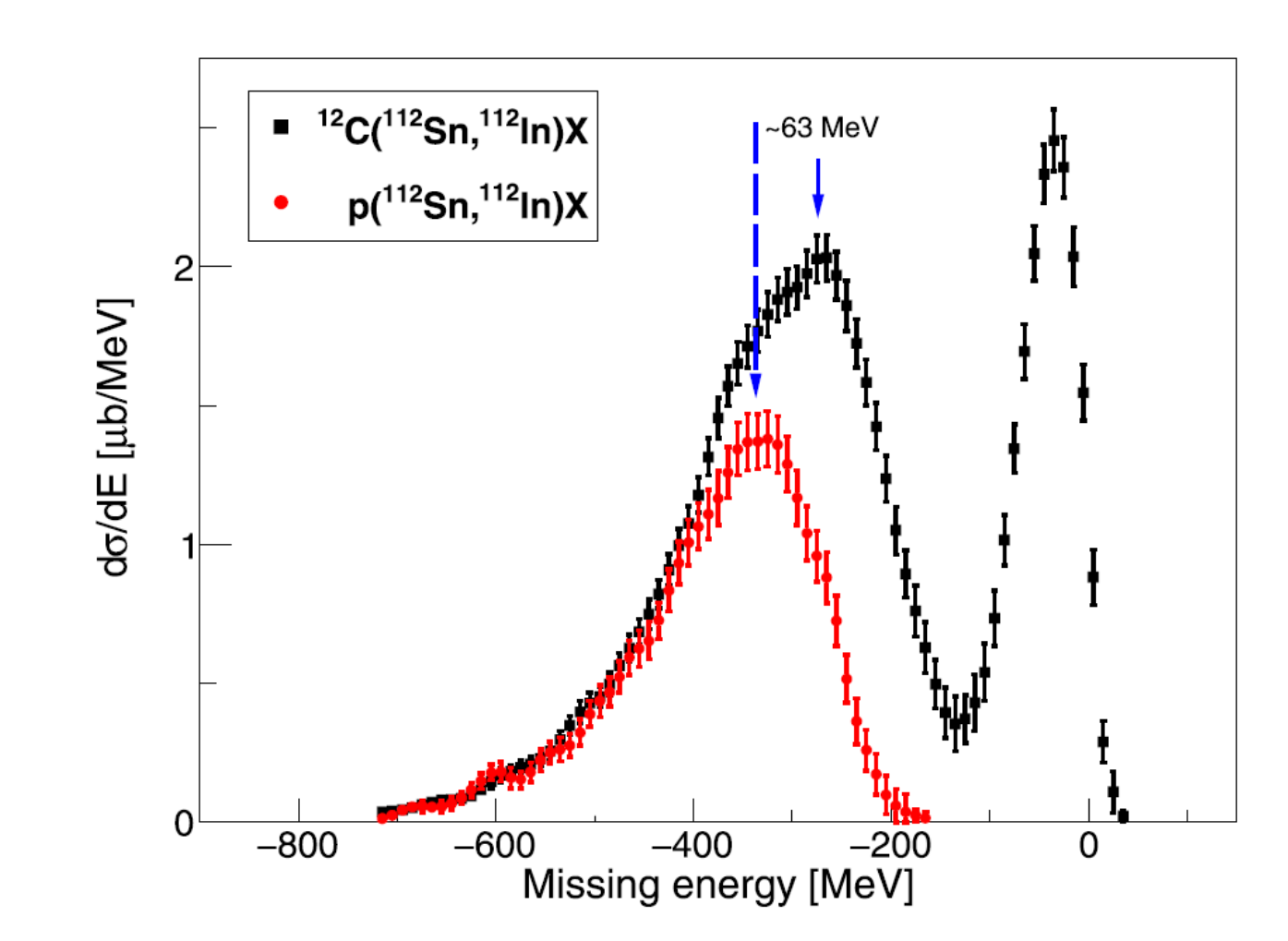}
\end{center}
\caption{Missing energy spectrum obtained in a $^{112}$Sn$\to{}{}^{112}$In reaction on a proton and a carbon target, in the latter case showing clearly separated quasi-elastic and resonance peaks. Note the shift in the resonance peak for the two targets (from Ref. \cite{Rodriguez-Sanchez:2020hfh}).}
\label{fig:MissE_112Sn}
\end{figure}

Grazing heavy ion reactions provide the proper conditions for spectroscopic investigations. Hence, peripheral reactions are the method of choice for large scale studies of various aspects of nuclear spectroscopy from quasi-.elastic (QE) over resonance (RE) excitations to deep-inelastic (DIS) scattering \cite{Lenske:2019cex,Cappuzzello:2022ton,Lenske:2024dsc,Lenske:2024mdc}. The existing FRS@GSI, previously and presently used for studies of exotic nuclei, is well suited for measurements of CC reactions, allowing even to use unstable beams  \cite{Rodriguez-Sanchez:2020hfh,Rodriguez-Sanchez:2021zik}.
Typical FRS-results for a $^{112}$Sn beam are shown Fig. \ref{fig:MissE_112Sn}, populating simultaneously $\tau_\pm$ ejectile channels $^{112}$Sb and $^{112}$In, respectively, while the target system is exposed to the complementary $\tau_\mp$ processes. The data show clearly the wealth of information gained by CC reactions at AGeV energies, namely reaching deeply into the RE region.

The presently available, highly valuable data are on inclusive cross sections, resolving spectroscopic details or decay channels. However,
the clearly resolved two-component structures are signatures of the change from nucleon $N'N^{-1}$ QE modes to $N^*N^{-1}$ RE excitations. The purpose of this work is to present updated theoretical methods which allow to describe the energy distributions of heavy ion CC reactions over large ranges of excitation energies, filtering out the global - and universal - features of QE and RE spectra.

In order to access the phenomena  expected in the experimentally covered energy regions, a configuration space of nucleons and - short-lived - nucleon resonances will be used. The theoretical discussion is centered around the first two excited nucleon states, the Delta resonance, $P_{33}(1232)$, and the so-called Roper resonance, $P_{11}(1440)$ \cite{PDG:2024cfk,Lenske:2025idu}.  The formalism, however, is open for an extensions of the spectrum of $N^*$  components.

while the Delta resonance - the first-ever observed excited sate of the nucleon \cite{Fermi:1952zz} - is since long a safely confirmed member of the baryon decuplet, the Roper resonance, observed as the second of kin by L.D: Roper \cite{Roper:1964zza}, is in fact still an especially controversial case which hardly can be assigned to a definite SU(3) multiplet. Since that state carries the same spin-isospin quantum numbers as the nucleon, the puzzling situation occurs that quark models prefer to interpret $P_{11}(1440)$ as a radial, i.e. compressional excitation of the nucleon. However, that is in conflict with hadron spectroscopy which assigns a strong nucleon-meson component to the same state, see the PDG compilation \cite{PDG:2024cfk} and  e.g. \cite{Shklyar:2014kra}. The latest experimental and theoretical results, summarized in \cite{Burkert:2025aze}, strongly favor the picture that $P_{11}(1440)$ is indeed a state composed of a $qqq$ core which is surrounded by a substantial meson cloud.

In this paper, the focus will be solely on modeling the nuclear CC response functions which were used in \cite{Rodriguez-Sanchez:2020hfh,Rodriguez-Sanchez:2021zik} to interpret the cross section data.  At the high beam-energies of the experiment,  ISI/FSI could be treated in Glauber theory. CC modes are of general interest, much beyond exploring reaction mechanisms and isovector nuclear spectroscopy. For example, spectra of that kind are playing a central role in neutrino-matter interactions as emphasized by the NuSTEC collaboration\cite{Alvarez-Ruso:2025oak}, being studied theoretically by similar methods as will be discussed in the forthcoming sections  \cite{Martini:2009uj,Martini:2010ex}. Transport-theoretical studies consider the same processes under slightly different aspects and with different theoretical methods \cite{Helgesson:1994mb,Helgesson:1998gs,Helgesson:1998gs,Larionov:2003av,Buss:2011mx}.

\begin{figure}
\begin{center}
\includegraphics[width=6cm]{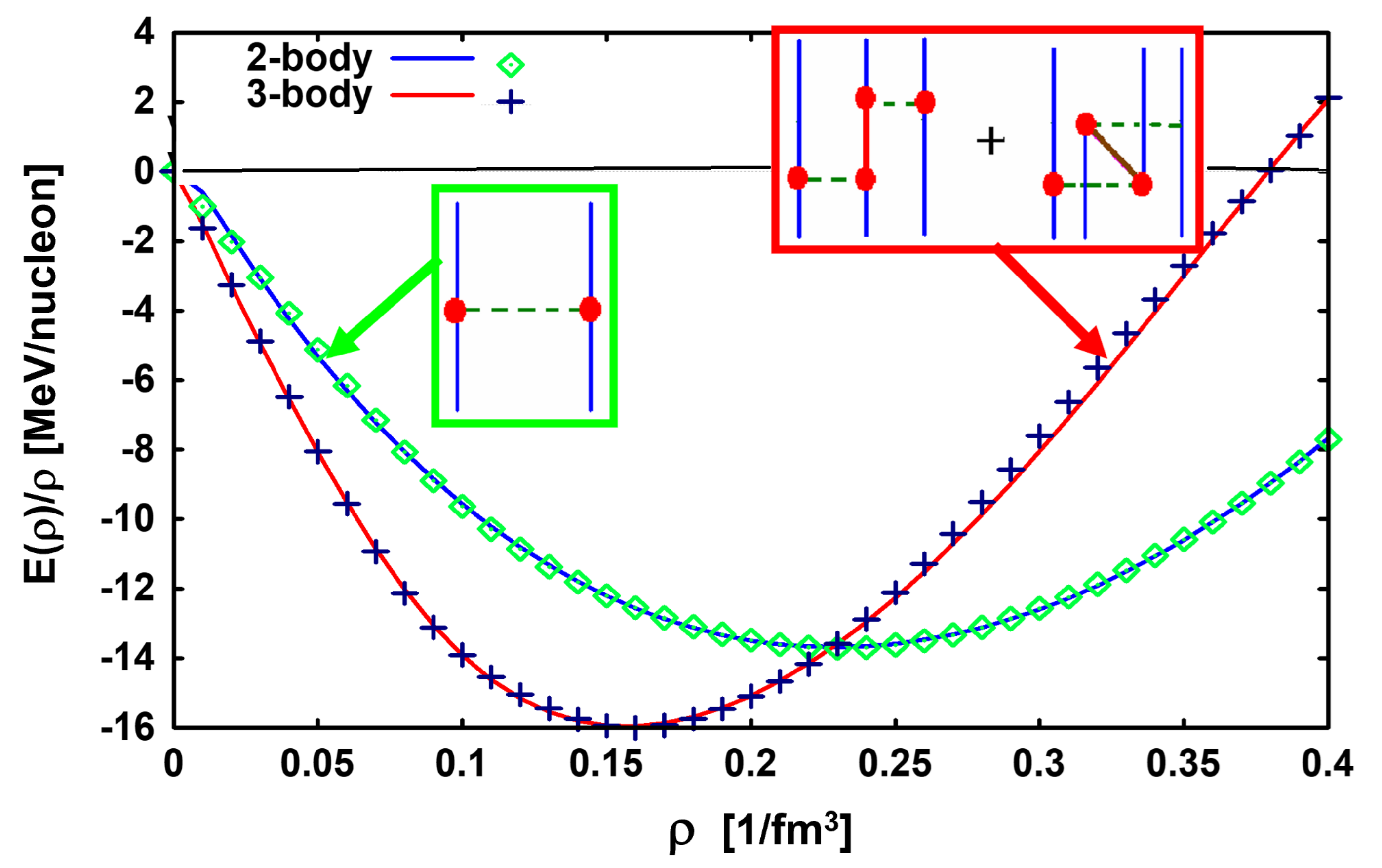}
\end{center}
\caption{Binding energy per nucleon of symmetric nuclear matter obtained with the GiEDF without (blue line/green symbols) and with many-body corrections (red line/purple symbols) of the Urbana model  UIX as used in \cite{Akmal:1998cf}. The  ladder and three-body resonance contributions are indicated. }
\label{fig:EoS}
\end{figure}

Investigations of the dynamical properties of a quantummechanical many-body system depend by obvious reasons on the proper description of the ground state relative to which dynamics are defined and are explored. In this respect approaches like those of \cite{Martini:2009uj,Martini:2010ex} are incomplete irrespective of their (relative) success and their importance for the theory of CC lepton-nucleus interactions. An intrinsically closed approach is obtained by energy density functional (EDF) theory and the methods of density functional theory (DFT). A nuclear EDF is a functional of the field operators of the active (massive) particles. In our case these are primarily protons and neutrons, defining the background medium, supplemented by nucleon resonances $N^*$ which by interaction with an external probe is enforced into $N'N^{-1}$ nucleon-nucleon  and $N^*N^{-1}$ resonance-nucleon particle-hole modes. The decisive theoretical step is how to combine the nuclear many-body environment and the phenomenology of short-lived baryon resonances. EDF approaches aim at a scheme for the systematic treatment of the various aspects of many-body dynamics from ground state properties to soft collective excitations and to hard collisional modes. EDF theory \emph{per se} does not attempt a so-called \emph{\emph{ab initio}} modelling  of nuclear systems but does not exclude such an approach \cite{Furnstahl:2019lue}, albeit it is unclear whether the power counting schemes developed under free space conditions is the same in high density nuclear matter.
A major advantage of the EDF approach is that
inconsistencies are avoided which may arise when incompatible interactions and theoretical methods are used.

The Giessen EDF (GiEDF) approach is based on microscopic descriptions of NN interactions. Interaction are derived from covariant Dirac-Brueckner G-matrices \cite{Lenske:1995,Fuchs:1995,Dejong:1998,Keil:1999hk,Hofmann:2000mc} and  non-covariant Hartree-Fock-Brueckner G-matrices \cite{Hofmann:1998,Lenske:2019ubp,Adamian:2021gnm}, both supplemented by density dependent (static) many-body correlations as practiced in the Urbana approach, see e.g. \cite{NS3}. The three-body additives are mainly done for infinite nuclear matter while adjustments to nuclear data are avoided.  In Fig. \ref{fig:EoS} the important interplay of two-body NN dynamics and many-body corrections of the three-body resonance type in the static limit are illustrated for GiEDF results on the binding energy per nucleon of infinite symmetric nuclear matter.

In the following sections the theory of nuclear charged current spectroscopy by response functions of nucleonic and resonance excitation modes will be summarized, thereby generalizing former versions \cite{Lenske:2018bvq,Lenske:2019cex} to an enlarged set of resonances. The aim is to describe nuclear CC response functions over large energy ranges up to the onset of deep-inelastic interactions in the GeV region. Results for spectra of $\tau_+$ and $\tau_-$ spin-scalar (Fermi-type) and spin-vector (Gamow-Teller-type) modes and the corresponding longitudinal and transversal spectra are presented. Formal details on propagators and transition operators ar explained in the appendices. The paper closes with a summary and outlook.

\section{Theory of $NN^{'-1}$ and $N^*N^{-1}$ Excitations}\label{sec:Theory}

\subsection{Energy Density Functionals and Interactions}
The proto-type of a purely phenomenological nuclear density functional is the widely and successfully used non-relativistic Skyrme model \cite{Vautherin:1971aw}, parameterized by kinetic energy densities and a set of contact interactions. A microscopically approach nased on two-body interactions was investigated in \cite{Negele:1972zp} and with the density matrix expansion a Skyrme-like EDF could indeed be derived - albeit suffering from the lack many-body interactions. On the covariant side, inspired by the early work of Walecka and Serot \cite{Walecka:1975ft,Serot:1984ey}, phenomenological Relativistic Mean-Field (RMF) models have become popular.

Under theoretical aspect, energy density functionals are the Hamiltonian density belonging to a field-\.theoretical Lagrangian. However, for investigation of many-body systems, defined by a pre-defined, fixed number of protons, neutrons, and eventually other baryons, it is of advantage to change coordinated from matter field-operators to invariant binomials of matter field-operators. Symmetries and conservation laws are conserved is the binaries are chosen such that they are related to the conserved quantum numbers of the system. A basic quantity of that kind is baryon number. The related Lorentz-invariant operator is the baryon four-current $j_\mu$ which defines the invariant number density of a system with A particles in its ground state $|A\ran$,  $\rho_A=\sqrt{\lan A|j_\mu j^\mu|A\ran}$ \cite{Lenske:2004} - as exploited in Hohenberg-Kohn \cite{Hohenberg:1964} and Kohn-Sham \cite{Kohn:1965}. Since $j_\mu=\sum_B \overline{\Psi}_B\gamma_\mu \Psi_B$ is a binomial of the baryon field operators, the related squared density operator $\widehat{\rho}=j_\mu j^\mu$ defines the mentioned change of coordinates.

Once the reference system is chosen, e.g. a nucleus $A(Z,N)$ or infinite nuclear matter composed of proton and neutron densities or neutrons star matter of certain composition, we may investigate the response of the system of infinitesimal variations $\delta\rho_B$ of the matter densities. That is achieved by a Taylor series expansion in the abstract functional density operator space around the expectation value $E_A(\rho_A)\equiv E(\rho_A)$, $\rho_A=\lan A|\widehat{\rho}|A\ran$ as investigated in Fermi-liquid theory \cite{Noziere:1964,Migdal:1967}. Up to second order one finds
\be
\mathcal{E}(\widehat{\rho})\approx E_A(\rho_A)+\sum_B\delta\rho_B U_B(\rho_A)+\frac{1}{2}\sum_{B,B'}\delta\rho_B\delta\rho_{B'}F_{BB'}
\ee
which is a harmonic approximation in (a high-dimensional) density space. Particle-particle interactions and pairing mean-fields are easily added. The generated terms are of physical relevance: $E_A(\rho_A)$ is the total mass of the reference state $A$, the latter playing the role of a vacuum state, the first variational derivatives describe the response of the system on removal or addition of particles of kind $B$, giving access to mean-field potentials (and energies) of the constituents $B$. $F_{BB'}$ are the restoring forces (Landau-Migdal parameters) of the system against number density variations which define the residual interactions by which dynamical processes are governed.

If $\mathcal{E}$ was defined with effective, density dependent in-medium two-body interactions $\mathcal{V}_{BB'}(\widehat{\rho})$, one finds
\be
U_B(\rho_A) = \sum_{B'}\mathcal{V}_{BB'}(\rho_A)\rho_{B'}+\frac{1}{2}\sum_{BB'}\rho_{B'}\rho_{B''}
\frac{\delta}{\delta \rho_B}_{|\rho_A}\mathcal{V}_{B'B''}(\rho_A).
\ee
The restoring forces are found as
\bea
F_{BB'}(\rho_A)&=&\mathcal{V}_{BB'}(\rho_A)+\sum_{B''}\rho_{B''}
\left( \frac{\delta}{\delta \rho_B}\mathcal{V}_{B'B''}(\rho_A) +
\frac{\delta}{\delta \rho_{B'}} \mathcal{V}_{BB''}(\rho_A) \right)\\
&+&\frac{1}{2}\sum_{B_1,B_2}\rho_{B_1}\rho_{B_2}
\frac{\delta^2}{\delta\rho_B\delta\rho_{B'}}_{|\rho_A}\mathcal{V}_{B_1B_2}(\rho_A) \nonumber.
\eea
The additional variational derivative terms are indispensable for the thermodynamical consistency of the theory \cite{Lenske:2004,Fuchs:1995} as expressed by the Hugenholtz-van Hove theorem \cite{Hugenholtz:1958}.

\subsection{Response Function Formalism for $N^*$ Configurations}\label{sec:RespF}

\begin{figure}
\begin{center}
\includegraphics[width=6cm]{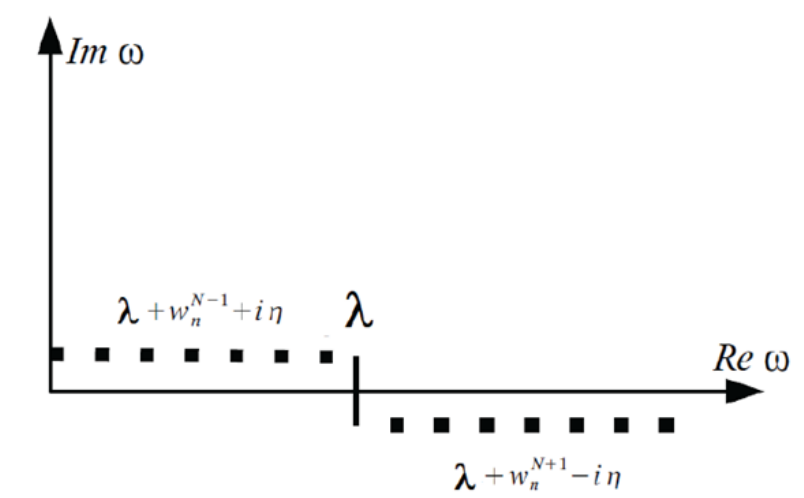}
\end{center}
\caption{Schematic illustration of spectrum relative to the chemical potential $\lambda$ of hole state and particle states which are located to the left and to the right of $\lambda$, respectively. In cold degenerate matter the chemical potential is given by the Fermi energy $\varepsilon_{Fq}$ of baryon type $q=p,n\ldots$. In momentum space the hole occupation numbers are given by the Heaviside distribution $n(\mathbf{k})=n^<(\mathbf{k})=\Theta(k^2_{Fq}-\mathbf{k}^2)$ while particle states are populated complementary, $n(\mathbf{k})=n(\mathbf{k})^>=1-n^<(\mathbf{k})=\Theta(\mathbf{k}^2-k^2_{Fq})$.  }
\label{fig:Spectrum}
\end{figure}

Formally, the approach appears as an one particle-one hole (1p1h) type Random Phase Approximation (RPA) which, however, incorporates mean-field self-energies as used in Fig.\ref{fig:EoS} and higher order contributions by nucleons and resonance in-medium self-energies. Thus, the basis states are dynamically dressed quasi-particles. The single hole and particle spectrum is illustrated in Fig.\ref{fig:Spectrum}. $N^*$ states include self-energies induced by their decay where in-medium modifications, as e.g. Pauli-blocking and pion absorption, are taken into account in analogy to the approach of Oset and Salcedo \cite{Oset:1985}. Analyticity is conserved only if both the real and the imaginary parts of the energy dependent dispersive polarization self-energies are used.

The creation of a resonances in a nucleus amounts to transform a nucleon into an excited intrinsic states. Thus, the nucleon is removed from the pre-existing Fermi-sea and a $N*N^{-1}$ configuration is created. That state is not an eigenstate of the many-body system but starts to interact with the background medium through residual interactions $V_{NN^*}$. The appropriate theoretical frame work for that process is given by the polarization propagator formalism \cite{Fetter:1971qmb}, also underlying, for example, the former approaches in \cite{Bertsch:1988xu,Helgesson:1994mb,Larionov:2003av}. 

\begin{figure}
\begin{center}
\includegraphics[width = 6cm]{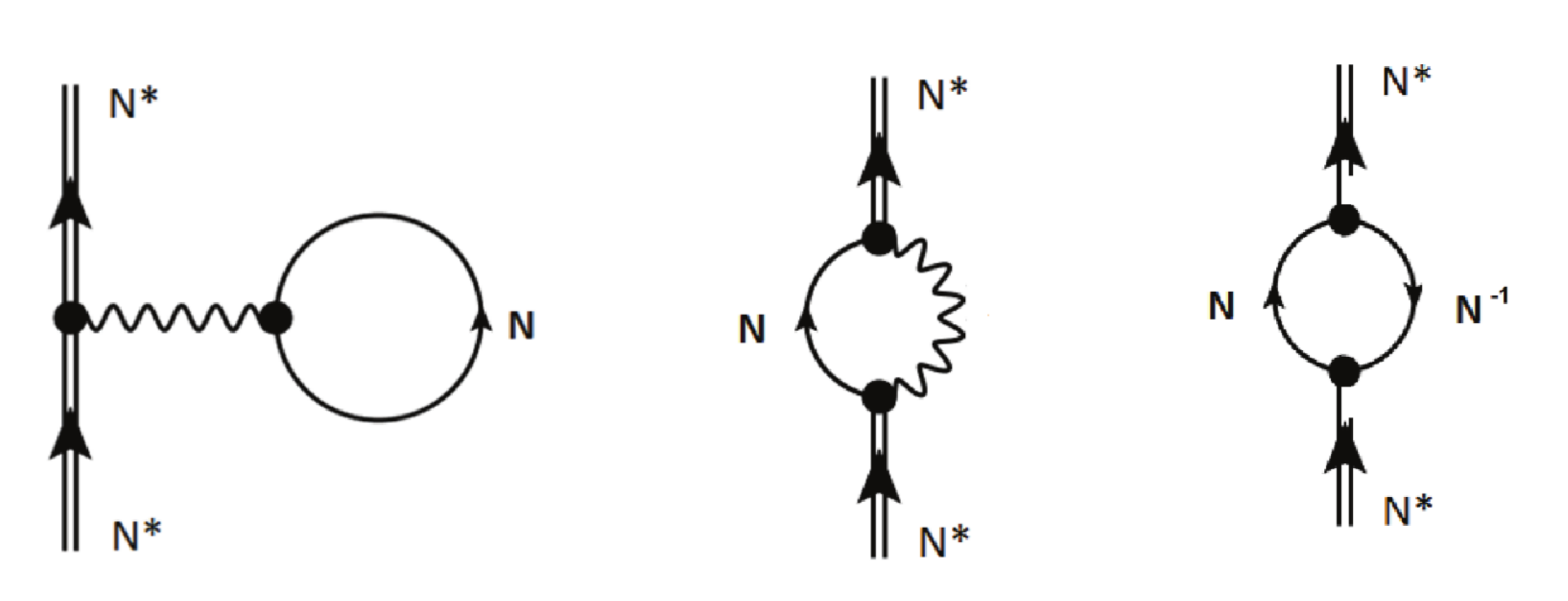}
\caption{In-medium interactions of a baryon resonance $N^*$ via the static mean-field (left) and the dispersive polarization self-energies (center) indicated here by the decay into intermediate nucleon-meson configurations. Moreover, in nuclear matter the coupling to $NN^{-1}$ excitations contributes a spreading width (right). Wavy lines indicate the exchange mesons $\pi,\eta,\sigma,\delta/a0(980),\rho,\omega$.}
\label{fig:Nstar_SelfE}
\end{center}
\end{figure}

For the non-interacting system the generalized particle-hole (ph) propagator is given by a block-diagonal matrix 
\be
\mathcal{G}^{(0)}(\omega ,\bf{q}) = \left( {\begin{array}{*{20}{c}}
{G_{N{N^{ - 1}}}^{(0)}(\omega ,\bf{q})}&0\\
0&{G_{N^* {N^{ - 1}}}^{(0)}(\omega ,\bf{q})}
\end{array}}\right)
\ee
where $G^{(0)}_{NN^{ - 1}}$ and $G^{(0)}_{N^* N^{ - 1}}$ describe the propagation of particle-hole (i.e. two-quasiparticle) $NN^{-1}$ and $N^* N^{-1}$ states, modified by the nuclear medium by mean--field interactions $U_{N,N^*}(\rho)$ and dispersive self-energies $\Sigma_{N,N^*}(\omega,\mathbf{q})$ - depending also on the density of the background medium - and decay self-energies for the unstable resonances, as displayed in Fig.\ref{fig:Nstar_SelfE}. Since $G^{(0)}_{N^* N^{ - 1}}$ represents the whole subspace of resonances, it is actually a diagonal sub-matrix of dimensionality equal to the total number of resonances.

These propagators are given by Lindhard functions for four-momentum transfer $k=(k_0,\mathbf{k})^T$, defined by integrations over the product of a hole propagator $G_N$ and a particle-state propagator $G_P$ \cite{FW:1971}
\be
\phi_{NP}(k)  =  i \int {d^4 p \over (2\pi)^4} G_N(p) G_P(p+k)~,  \label{LindH}
\ee
Depending on the kind of isospin mode,  $G_N(p)$  is a neutron or proton hole propagator while $G_{P}(p)$ is either the particle propagator of the complementary type of nucleon or a
resonance propagator. The formulation is oriented on the (non-relativistic) many-body framework of Ref.\cite{Fetter:1971qmb}.
The formalism is of general character and applicable to any isospin and spin-parity states, see e.g. the recent study of $\omega$-nucleus interactions \cite{Lenske:2023mis,Lenske:2024wjp}.

The propagator of proton and neutrons (in spectroscopic notation $P_{11}(940)$ \cite{PDG:2024cfk}) is
\be\label{nucprop}
G_N(p)  =  \frac{1}{ p^0 - \varepsilon({\bf p})-\Sigma_N(p^2) + i0} +
2 \pi i\, n({\bf p}) \delta( p^0 - \varepsilon({\bf p})-Re[\Sigma_N(p^2)] )~.
\ee
where the second term accounts for the  contributions of the Fermi-sea, occupied by the probabilities $n(p)$ at four-momentum $p=(p_0,\mathbf{p})^T$.
The propagator of  nucleon-like $P_{11}$ resonances is of the same structure but with $n(p)=0$. Single particle energies including the static mean-field self-energies are denoted by $\varepsilon_{N}$

Delta-resonances and resonances of higher spin obey their specific covariant wave equations and propagators which where studied e.g. in
\cite{Shklyar:2008kt,Shklyar:2009cx}. For the present purpose, Delta propagation is described by a simplified Rarita-Schwinger propagator which non-relativistically is
\be\label{eq:delprop}
G^{\mu\nu}_\Delta(p)  =
\frac{ 1}{ p^0 - \varepsilon_\Delta({\bf p})
  -\Sigma_\Delta(p^2) +i0}\delta^{\mu\nu}~.
\ee

After the contour integration over $p_0$ the nucleon particle-nucleon hole Lindhard function attains the form~:
\bea
&&\phi_N(k) =-\\
&&  \int {d^3 p \over (2\pi)^3} \left(
   { n({\bf p}+{\bf k})(1-n({\bf p})) \over
     \varepsilon^*({\bf p}+{\bf k}) - k^0 - \varepsilon^*({\bf p}) + i0 }
+  { n({\bf p})(1-n({\bf p}+{\bf k})) \over
     \varepsilon^*({\bf p}) + k^0 - \varepsilon^*({\bf p}+{\bf k}) + i0 }
                                           \right)~,\nonumber
                                                      \label{LindN1},
\eea
being summed over spin projections $s,s'$. $P_{11}$ resonances are described accordingly, but with $n_{s'}=0$-
Delta particle-nucleon hole Lindhard function becomes:
\begin{equation}
  \phi_\Delta(\pm k) = - \int {\frac{d^3 p}{ (2\pi)^3}
   { n({\bf p}) \over \varepsilon^*_N({\bf p}) - \varepsilon^*_\Delta({\bf p} \pm {\bf k}) \pm k^0
      }}~.      \label{LindD1}
\end{equation}
The in-medium energies $\varepsilon^*_{N,\Delta}({\bf p})=\varepsilon_N({\bf p})+\Sigma_{N,\Delta}(p)$ include dynamical self-energies. Since the single particle self-energies were chosen as spin-independent, the propagators apply equally to al spin-states of nucleons and resonances. In appendix \ref{app:SpinIso} the properties of Lindhard function in ANM are discussed in detail.

The residual $NN^{-1}$ and $N^* N^{-1}$ interactions are contained b in
\be
\mathcal{V} = \left( {\begin{array}{*{20}{c}}
{{V_{NN}}}&{{V_{NN^* }}}\\
{{V_{N^* N}}}&{{V_{N^* N^* }}}
\end{array}} \right).
\ee
The Green function of the interacting system is given by the Dyson equation of the 4-point function
\be
\mathcal{G}(\omega,\mathbf{q})=
\mathcal{G}^{(0)}(\omega,\mathbf{q})+\mathcal{G}^{(0)}(\omega,\mathbf{q})\mathcal{V}\mathcal{G}(w,\mathbf{q}).
\ee
Since polarization self-energies induced by the coupling to higher order configurations are included, the approach is in fact a projection to the one particle-one hole sector, corresponding to an extended Random Phase Approximation (RPA) of nucleons and resonances.

\begin{figure}
\begin{center}
\includegraphics[width = 7cm]{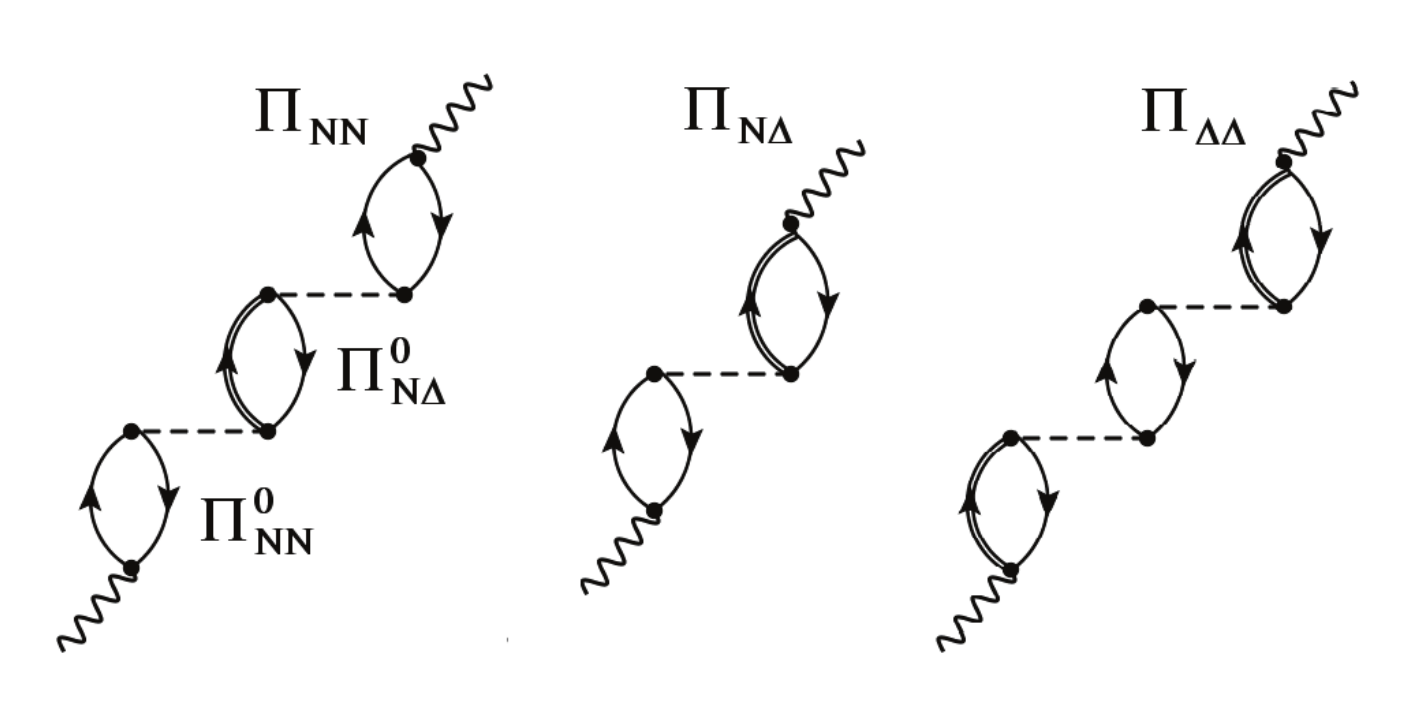}
\caption{The RPA polarization propagator. The $N^{-1}N\to N^{-1}N$ (left), the mixed $N^{-1}N\to N^{-1}\Delta$ and the $N^{-1}\Delta\to N^{-1}\Delta$ components are displayed. Also the bare particle-hole type propagators are indicated. External fields are shown by wavy lines, the residual interactions are denoted by dashed lines. Only part of the infinite RPA series is shown.}
\label{fig:RPA_Pol}
\end{center}
\end{figure}

\subsection{Response Function Formalism}
The coherent response of the $A=Z+N$ nucleon system with ground sate $|A\ran$ to an external probe with operator $\mathcal{O}_a(\mathbf{q})\sim e^{i\mathbf{q}\cdot \mathbf{r}} \bm{\sigma}^{S_a}\bm{\tau}^{T_a}$ where $a=(S_a,T_a)$ denotes scalar and vector spin ($S_a=0,1$), isospin ($T_a=0,1$) and momentum ($\mathbf{q}$) transfer, the via ($\sim e^{i\mathbf{q}\cdot \mathbf{r}}$), is described by the polarization propagators of the non-interacting system
\be
\Pi^{(0)}_{ab}(\omega,\mathbf{q})=\lan A|\mathcal{O}^\dag_b \mathcal{G}^{(0)}(\omega,\mathbf{q}) \mathcal{O}_a |A\ran.
\ee
The polarization propagators of the interacting system for the chosen set of operators is
\be
\Pi_{ab}(\omega,\mathbf{q})=\lan A|\mathcal{O}^\dag_b \mathcal{G}(\omega,\mathbf{q}) \mathcal{O}_a |A\ran.
\ee
obey the Dyson equation:
\bea\label{eq:PolTens}
&&\left( {\begin{array}{*{20}{c}}
{\Pi _{NN}^{}}&{\Pi _{N\Delta }^{}}\\
{\Pi _{N* N}^{}}&{\Pi _{N* N* }^{}}
\end{array}} \right) =
\left( {\begin{array}{*{20}{c}}
{\Pi _{NN}^{(0)}}& 0 \\
 0 &{\Pi _{N* N* }^{(0)}}
\end{array}} \right)\\
&&+
\left( {\begin{array}{*{20}{c}}
{\Pi _{NN}^{(0)}}& 0 \\
 0 &{\Pi _{N* N* }^{(0)}}
\end{array}} \right)\left( {\begin{array}{*{20}{c}}
{{V_{NN}}}&{{V_{NN* }}}\\
{{V_{N* N}}}&{{V_{N* N* }}}
\end{array}} \right)\left( {\begin{array}{*{20}{c}}
{\Pi _{NN}^{}}&{\Pi _{NN* }^{}}\\
{\Pi _{N* N}^{}}&{\Pi _{N* N* }^{}}
\end{array}} \right)\nonumber
\eea
which is illustrated diagrammatically in Fig.\ref{fig:RPA_Pol}.

Once the polarization propagator is known, observables are easily calculated. The response functions. i.e. the spectral distribution of transition strengths, are defined by
\be
S_{ab}(\omega,\mathbf{q})=-\frac{1}{\pi}Im\left [ \Pi_{ab}(\omega,\mathbf{q}) \right].
\ee
Summations over spin quantum numbers are implicit.

\section{CC Response Functions of Asymmetric Nuclear Matter}
\subsection{Lindhard Function in Asymmetric Matter}
The translation invariance of infinite matter simplifies calculations significantly. Single particle wave functions are given by plane waves times a spin-isospin wave function. The matrix elements of $\mathcal{V}=\mathcal{V}(q)$ are then given by the Fourier-Bessel transforms of the respective particle-hole two-body interactions and spin-isospin matrix elements. The same is true for the matrix elements of the one-body operators $\mathcal{O}_{a,b}$.

Investigations of asymmetric infinite nuclear matter (ANM) have to account for the spontaneously broken isospin symmetry by the imposed difference in proton and neutron content. That is taken into account by introducing separate proton and neutron propagators which are distinguished by different ground state occupation numbers as defined by the respective Fermi-momenta $k_{Fp}\neq k_{Fn}$ - or, in a more general context,  the respective chemical potential $\lambda_p\neq \lambda_n$. As a consequence, also the particle-hole propagators double in number by requiring not only $pn^ {-1}$ and $np^ {-1}$ 4-point functions but also
$N^*n^ {-1}$ and $N^*p^ {-1}$.

In leading order, i.e. mean-field approximation, the nucleon ground state occupation numbers are given by Heaviside distributions, $n_q(p)=\Theta(k_{Fq}-|\mathbf{p}|)$ for $q=p,n$ and the respective Fermi momentum $k_{Fq}$.
Hence,
the $pn^{-1}$ Lindhard functions of $\tau_+$ modes are
\be
\phi^{(+)}_N(\omega,\mathbf{q})=\phi_{np}(\omega+i\eta,\mathbf{q})-\widetilde{\phi}_{pn}(\omega-i\eta,\mathbf{q})
\ee
where $\eta \to 0+$. With $w=\omega+i\eta$ the so-called \emph{time-forward} amplitude is defined by
\be
\phi_{pn}(w,\mathbf{q})=-\int \frac{d^3k}{(2\pi)^3}
\frac{\Theta(k_{Fp}-k)\Theta(|\mathbf{k}+\mathbf{q}|-k_{Fn})}
{w-\varepsilon^*_n(\mathbf{k}+\mathbf{q})+\varepsilon^*_p(\mathbf{k})-\Sigma_{np}(\omega,\mathbf{q})}
\ee
and the \emph{time-backward} amplitudes are
\be
\widetilde{\phi}_{pn}(w,\mathbf{q})=-\int \frac{d^3k}{(2\pi)^3}
\frac{\Theta(k_{Fp}-|\mathbf{k}+\mathbf{q}|)\Theta(|\mathbf{k}|-k_{Fn})}
{-w^*-\varepsilon^*_p(\mathbf{k}+\mathbf{q})+\varepsilon^*_n(\mathbf{k})-\Sigma_{pn}(\omega,\mathbf{q})}.
\ee
In the resonance sector the backward amplitudes do not appear. Correspondingly, the $N^*N^{-1}$ Lindhard functions of the nucleon-resonance modes are changed into the set of distinct $N^*p^{-1}$ and $N^*n^{-1}$ Lindhard functions. Further details on the structure and properties of Lindhard functions are found in Appendix \ref{app:SpinIso}.

\subsection{CC Response of Asymmetric Infinite Nuclear Matter}

In the concrete case of numerical calculations on the $\tau_\pm$ modes of ANM the configuration space was built of protons, neutrons, the $P_{11}(1440)$ Roper- and the $P_{33}(1232)$ Delta-resonance. The $\tau_+$ modes then involve
$pn^{-1}$ quasi-elastic excitations. In the inelastic resonance region one finds the corresponding Roper mode $P^+_{11}n^{-1}$ and two Delta-modes, $P^+_{33}n^{-1}$ and also $P^{++}_{33}p^{-1}$. The Roper- and Delta-modes contribute also to the $\tau_-$ sector where in addition to $np^{-1}$ excitations the spectra contain the Roper-mode
$P^0_{11}p^{-1}$ and the two Delta-modes, $P^0_{33}p^{-1}$ and also $P^{-}_{33}n^{-1}$. Hence, in ANM the main contributions to $\tau_+$ and the $\tau_-$ excitations are built on different nucleon Fermi seas which, however, become mixed by Delta-modes.
In symmetric nuclear matter (SNM) the same kind of spectral features are found but since isospin symmetry is restored in SNM the $\tau_\pm$ response functions will be equal.
The same composition and features will be found in CC spectra of finite nuclei. However, electromagnetic effects, will inhibit even in $N=Z$ nuclei the full restoration of isospin symmetry.

\begin{figure}
\begin{center}
\includegraphics[width = 12cm]{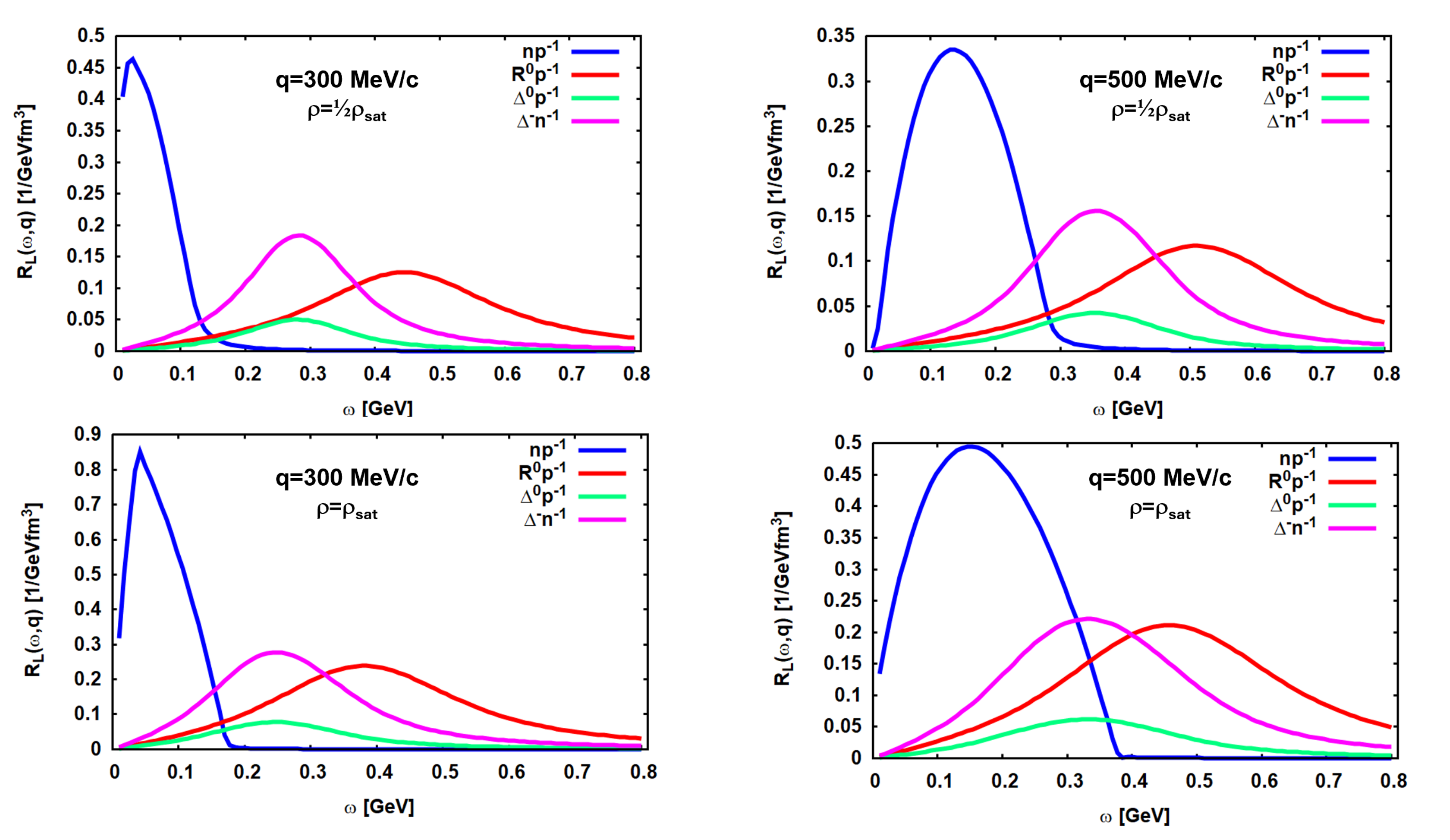}
\caption{Longitudinal RPA response functions for the operator $\bm{\sigma}\cdot \mathbf{p} \tau_-$ in asymmetric nuclear matter with a proton content of 40\%. In the upper and lower row, results at total density $\rho=\frac{1}{2}\rho_{sat}$ and $\rho=\rho_{sat}=0.16fm^{-3}$ are shown. In the left column and the right column response functions at momentum transfer $q=300$~MeV/c and $q=500$~MeV/c, respectively, are displayed.
at  momentum transfer $q=300$~MeV/c and in the right column results at $q=500$~MeV/c are shown. }
\label{fig:ANM}
\end{center}
\end{figure}

In Fig.\ref{fig:ANM}, response functions of ANM with 10\% neutron excess ($\zeta=Z/A=0.4$) for the $\tau_-$  member of the iso-triplet of longitudinal CC spin operators $\mathcal{O}^{(\bm{\tau})}_L=\bm{\sigma}\cdot \mathbf{p} \tau_\pm$ are displayed, illustrating the density dependence and the dependence on the three-momentum transfer. With increasing  density an upscaling of the spectral distribution is seen, reflecting the enlargement of the Fermi spheres with increasing density. The momentum dependence is significantly stronger as seen by the broadening of the quasi-elastic component and the gain in strength and width of the Delta- and Roper-contributions.

The (pseudo-scalar) longitudinal operator is of special interest because it is the pion-nucleon vertex. The complementary transversal iso-triplet of (vector) operators $\mathbf{\mathcal{O}}^{(\bm{\tau})}_T=(\bm{\sigma}\times \mathbf{p}) \bm{\tau}$, describing the $\bm{\rho} N$ isovector-spin vector vertices. Other operators of interest - also for weak CC processes - are the spin-scalar (Fermi) operate $F^{(\bm{\tau})}=\bm{\tau}$ and the spin-vector (Gamow-Teller) operator $\mathbf{G}^{(\bm{\tau})}=\bm{\sigma}\bm{\tau}$, both promoting NC and CC transitions. In appendix \ref{app:2ndQ} (one-body) transition operators are represented in second quantization.

\subsection{In-Medium $N^*$ Spectral Distributions}

\begin{figure}
\begin{center}
\includegraphics[width = 8cm]{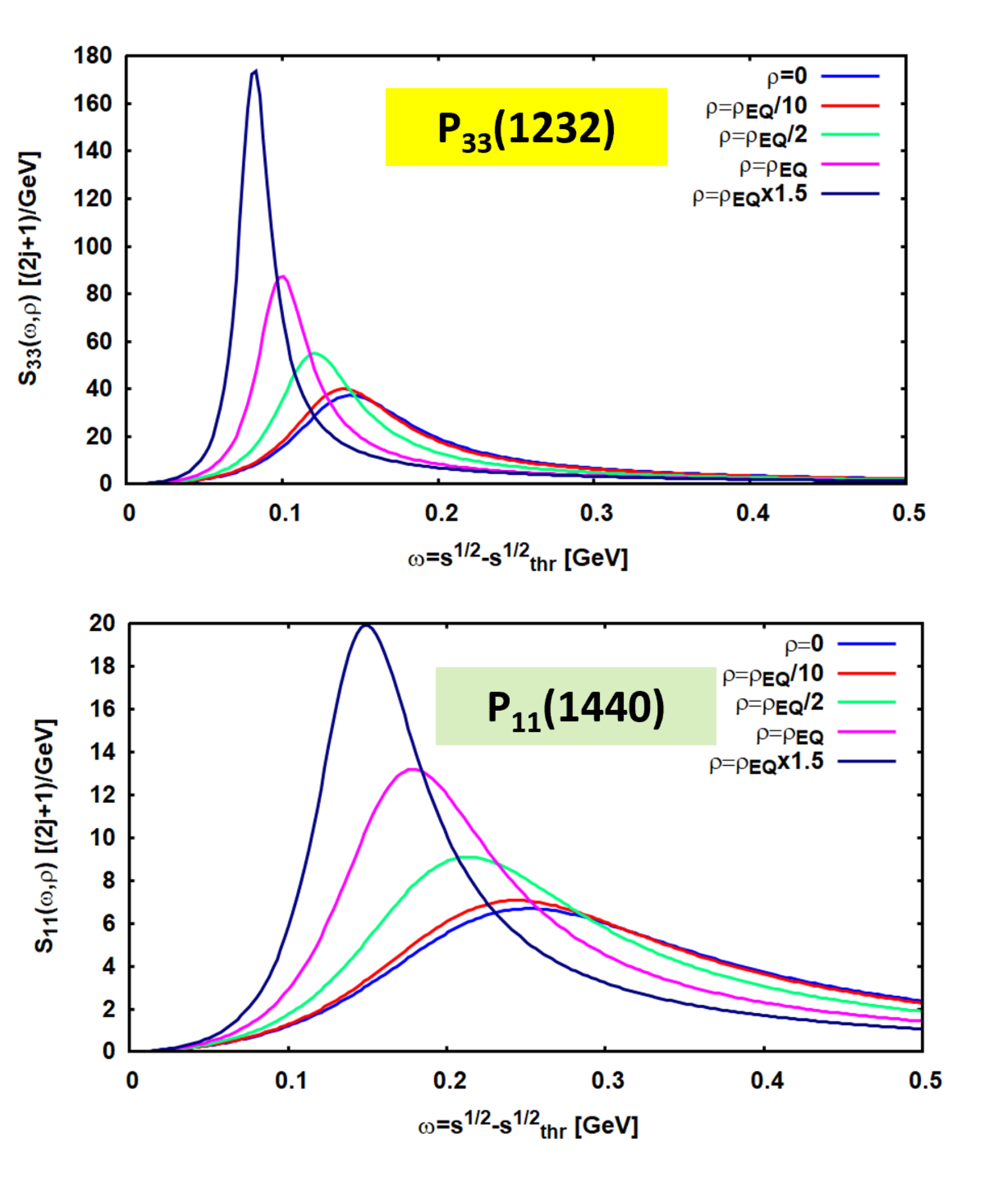}
\caption{Spectral distribution of the P-wave resonances in free space (blue lines) and in infinite nuclear matter at a selection of densities as indicated. }
\label{fig:Width}
\end{center}
\end{figure}

The density and momentum dependencies are largely driven by the combined action of static mean-field and dynamical in-medium self-energies. Nucleon and resonance propagators include
mean-field potentials, effective kinetic masses and polarization self-energies from the coupling to particle-hole modes of the background medium, for resonances being supplemented by the decay self-energies. The latter are corrected for
Pauli-Blocking and in-medium pion absorption, using an updated version of the Oset-Salcedo model \cite{Oset:1987re}.
As seen Fig.\ref{fig:Width}, the net effect is a shift of the peak positions towards lower energies, caused by mean-field attraction and the real part of the self-energies, accompanied by a reduction in width. The relocation reduces the phase space available for decay and Pauli-blocking and increased pion absorption act into the same direction of reducing the resonance decay width in nuclear matter.

\section{Charged Current Response Functions in Finite Nuclei}
\subsection{Local Density Approximation}
An elegant method for investigating the global spectral properties of finite nuclei, focused on universal features, is the Local Density Approximation (LDA). The  LDA amounts to use infinite matter response functions for ANM of the same asymmetry as the nucleus $A(Z,N)$ under scrutiny, $\zeta=Z/A$, but using local Fermi momenta $k_{Fq}\to k_{Fq}(\mathbf{r})=(3\pi^2\rho_q(\mathbf{r}))^\frac{1}{3}$ where for $q=p,n$ $\rho_q(\mathbf{r})$ is the (self-consistent HF/HFB) ground state density, preferentially obtained by the same interaction model as used in the ANM calculations. That approach was used in the afore mentioned spectral studies for the FRS experiments \cite{Rodriguez-Sanchez:2020hfh,Rodriguez-Sanchez:2021zik}.
The response functions of the nucleus $A$ are obtained by integration of they LDA spectral distribution
\be
S^{(A)}_{ab}(\omega,\mathbf{q})=-\frac{1}{\pi}\int d^3r
Im\left[\Pi^{(LDA)}_{ab}(\omega,\mathbf{q}|\rho_A(\mathbf{r})) \right]
\ee

\begin{figure}
\begin{center}
\includegraphics[width = 12cm]{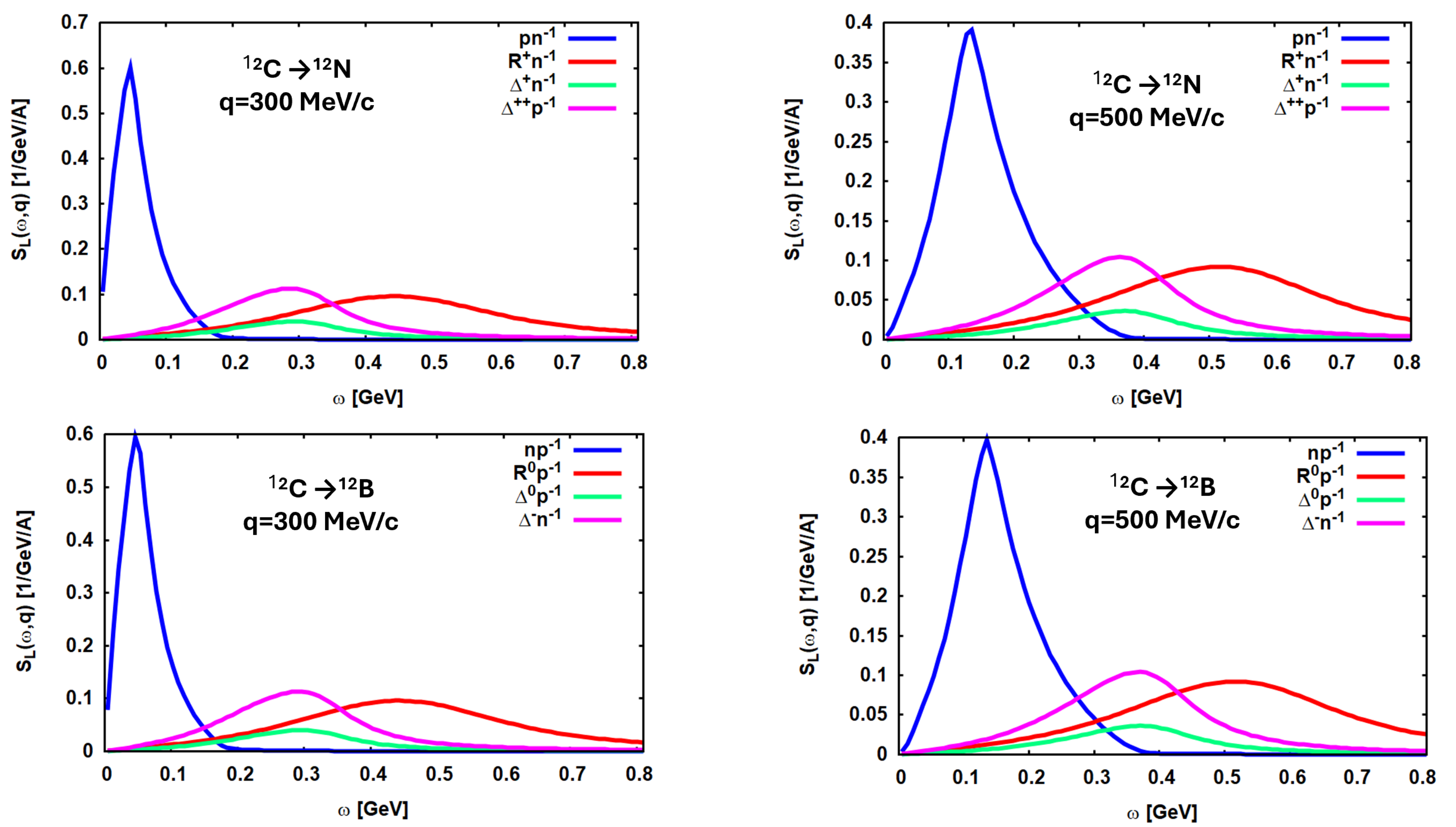}
\caption{Longitudinal spin-isospin response functions per nucleon for $^{12}$C$\to {}^{12}$N (upper row) and for
$^{12}$C$\to {}^{12}$B (lower row). In the left column, results
at  momentum transfer $q=300$~MeV/c and in the right column results at $q=500$~MeV/c are shown. }
\label{fig:LRPA_C}
\end{center}
\end{figure}

\begin{figure}
\begin{center}
\includegraphics[width = 12cm]{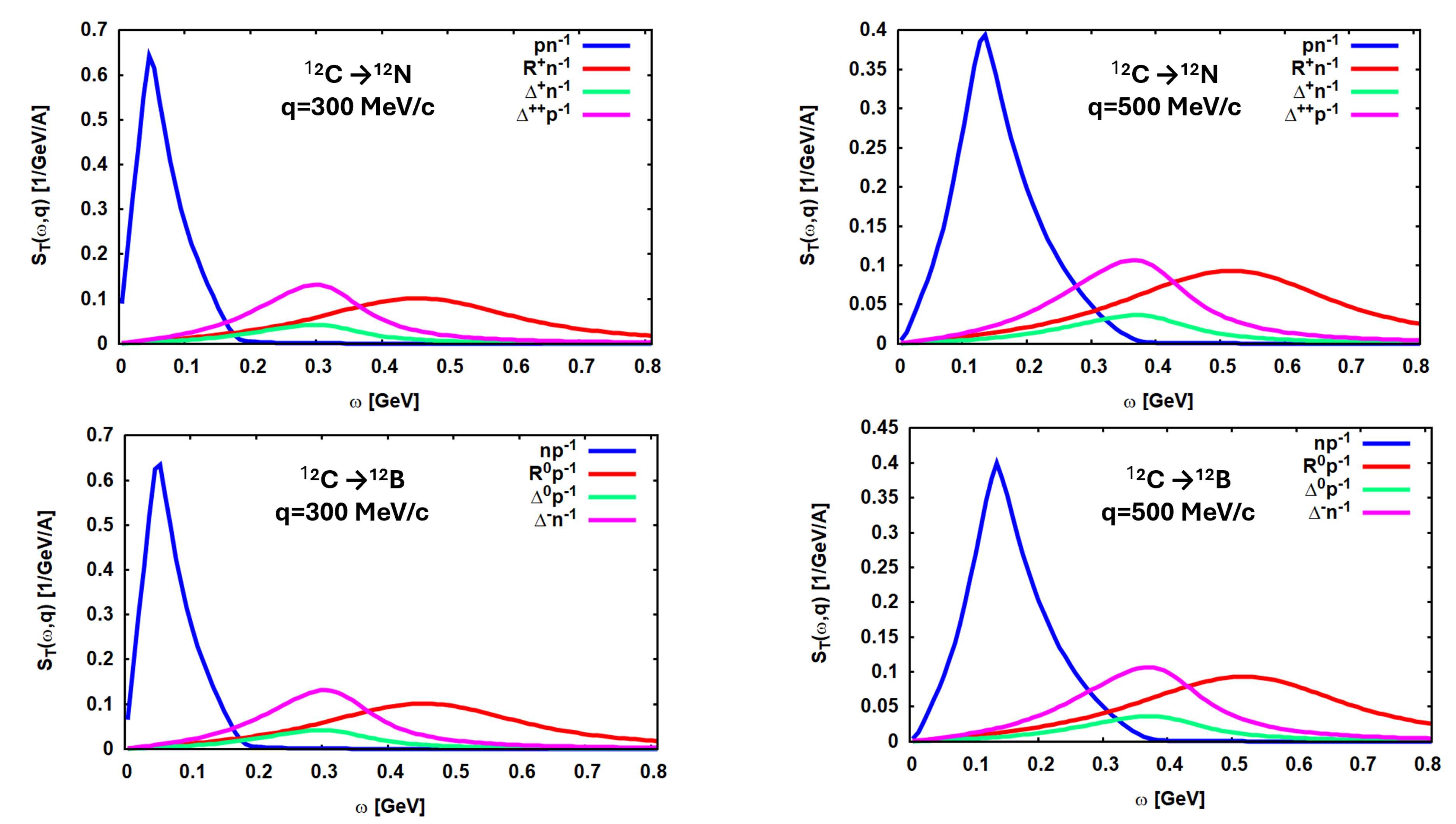}
\caption{Transversal spin-isospin response functions per nucleon for $^{12}$C$\to {}^{12}$N (upper row) and for
$^{12}$C$\to {}^{12}$B (lower row). In the left column, results
at  momentum transfer $q=300$~MeV/c and in the right column results at $q=500$~MeV/c are shown. }
\label{fig:TRPA_C}
\end{center}
\end{figure}

In Fig,\ref{fig:LRPA_C} and Fig.\ref{fig:TRPA_C} the $\tau_\pm$ longitudinal and transversal response functions for $^{12}$C are shown. The momentum dependence shows the same features as in ANM, although $^{12}$C is a $N=Z$ nucleus with asymmetry $\zeta=0.5$. Aside from minor differences in detail, overall  the longitudinal and transversal response functions are rather similar. Noteworthy is the string contribution of the Roper resonance. That nucleon partner state plays a prominent role in the CC response.

\begin{figure}
\begin{center}
\includegraphics[width = 12cm]{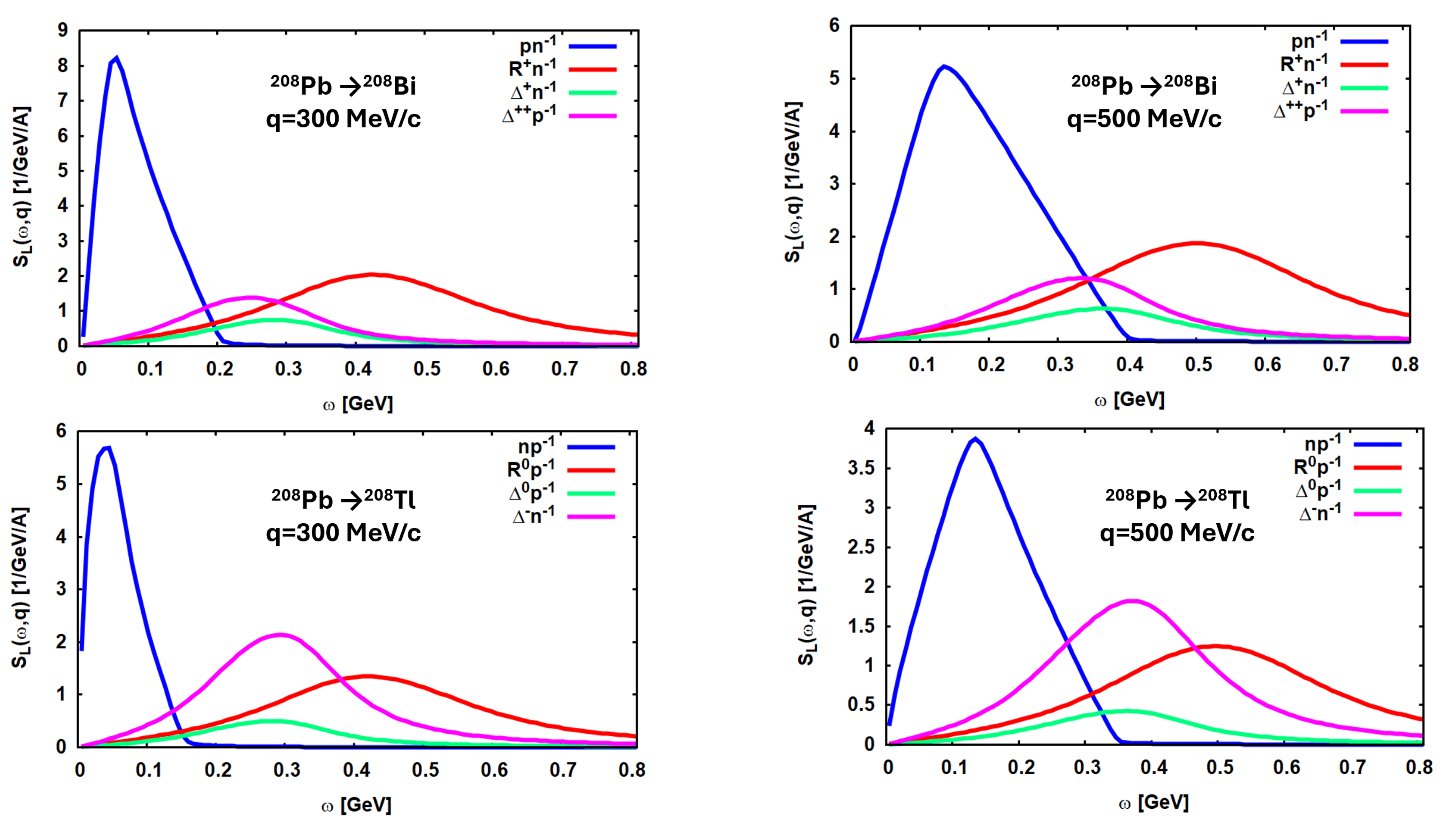}
\caption{Longitudinal spin-isospin response functions per nucleon for $^{208}$Pb$\to {}^{208}$Bi (upper row) and for
$^{208}$Pb$\to {}^{208}$Tl (lower row). In the left column, results
at  momentum transfer $q=300$~MeV/c and in the right column results at $q=500$~MeV/c are shown. }
\label{fig:LRPA_Pb}
\end{center}
\end{figure}

\begin{figure}
\begin{center}
\includegraphics[width = 12cm]{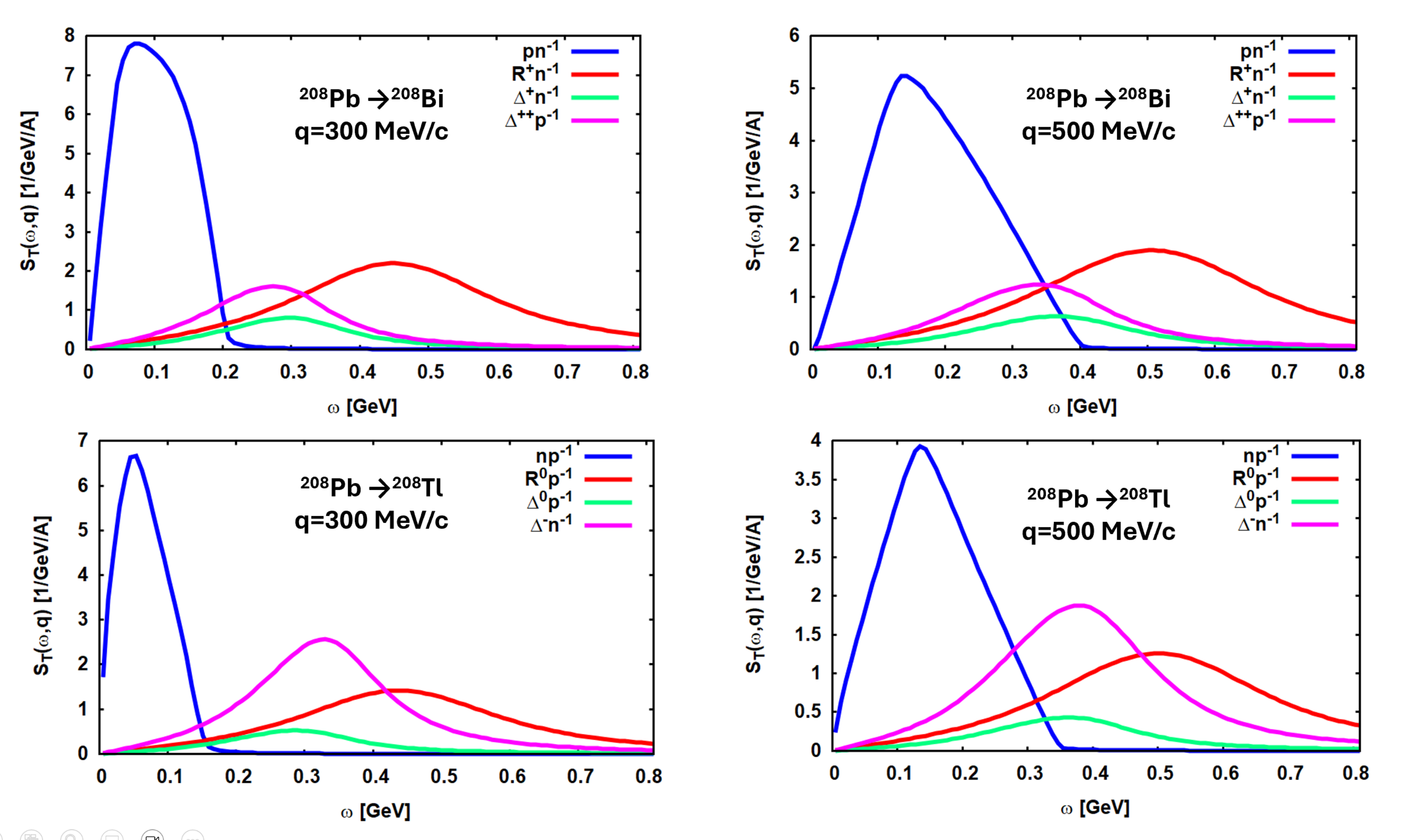}
\caption{Transversal spin-isospin response functions per nucleon for $^{208}$Pb$\to {}^{208}$Bi (upper row) and for
$^{208}$Pb$\to {}^{208}$Tl (lower row). In the left column, results
at  momentum transfer $q=300$~MeV/c and in the right column results at $q=500$~MeV/c are shown. }
\label{fig:TRPA_Pb}
\end{center}
\end{figure}

The features found in ANM and $^{12}$C are also visible the $\tau_\pm$ longitudinal and transversal response functions in $^{208}$Pb, shown in Fig.\ref{fig:LRPA_Pb} and Fig.\ref{fig:TRPA_Pb}, respectively. The that heavy nucleus, the resonance components appear to be enhanced. The Roper resonance component is especially enhanced  in the
$^{208}$Pb$\to {}^{208}$Bi $\tau_+$ case which largely is caused by the optimal overlap of the isospin wave functions within ind $P_{11}$ multiplet..

\subsection{Comparison to Data}
As mentioned before, the theoretical studies were motivated by experiments investigating peripheral CC heavy ion reactions at relativistic beam energies. Details on the experiments, the measurements, and data reduction together with detailed descriptions of the treatment of initial and final state interactions by Glauber theory are found in \cite{Rodriguez-Sanchez:2020hfh,Rodriguez-Sanchez:2021zik}.

\begin{figure}
\begin{center}
\includegraphics[width = 12cm]{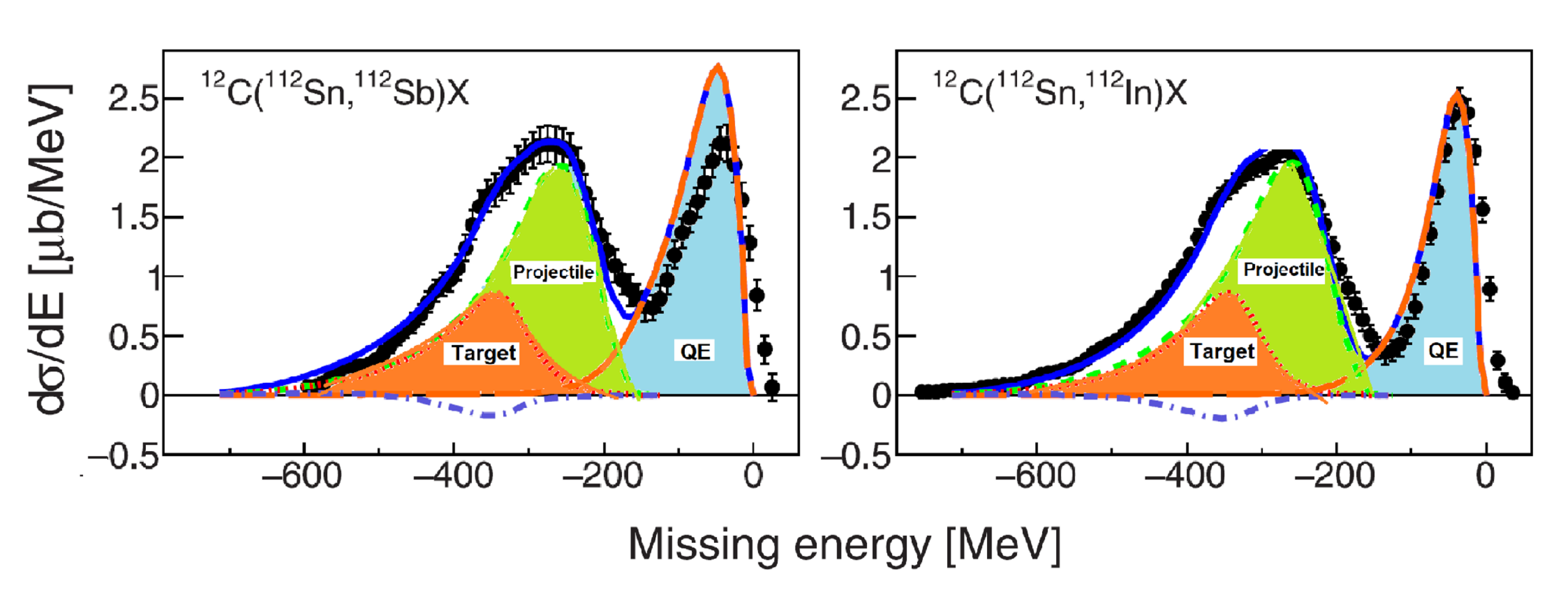}
\caption{Energy distributions of CC reactions of a $^{112}$Sn beam at 1AGeV on a $^{12}$C target. On the left, the energy-differential cross section recorded for $^{112}$Sb ejectiles and a X=$^{12}$B-like residual system constrained only by baryon number $A_T=12$ and total charge $Z_-=Z_T-1=5$. On the right, the energy distribution for the complementary exit channel $^{112}$In, accompanied by a X=$^{12}$N-like residual target system with $Z_+=Z_T+1$ electric charges. is displayed. The partial contributions of projectile and target excitation are indicated. }
\label{fig:dsde}
\end{center}
\end{figure}

As an illustration of the - at first sight surprising - success of the combined Glauber-plus-EDF and N$^*$RPA LDA approach the energy-differential cross sections for CC reactions of a $^{112}$Sn beam at 1AGeV on a $^{12}$C target are shown in Fig.\ref{fig:dsde}. Inclusive cross section results for the $\tau_+$-exit channel $^{112}$Sb and the complementary $\tau_-$-exit channel $^{112}$In are displayed. The related target-like $\tau_\mp$ residual systems are not ´resolved. They are constrained only by the total baryon number $A_T=12$ and the respective residual charge number $Z_-=Z_T -1=5$ and $Z_+=Z_T +1=7$, hence will include an unresolved number of mesons, leptons, and photons.

\section{Summary and Outlook}\label{sec:SumOut}

Nuclear CC modes are of large importance for understanding nuclear dynamics in the isospin sector. While in the long history of studies with light ion charge exchange reactions the low-energy quasi-elastic modes as the spin-scalar  Fermi isobaric analog resonance (IAR) and  spin-vector Gamow-Teller resonance (GTR) were in the center of interest, relativistic heavy ions beams have opened the window to large scale, systematic spectroscopic studies over energy regions covering the full range of quasi-elastic CC modes and penetrating deep into the sector of resonance excitations.

In the this paper, an EDF-based theoretical approach has been presented which allows to describe the spectrum of involved nuclei
microscopically. Mean-field dynamics and residual interaction were derived by first and second variation from the same EDF thus guaranteeing thermodynamical and mechanical consistency between the static and the in-medium dynamical parts of the nucleon resonance self-energies. Hence, $N$ and $N^*$ propagators are compatible which will reduce the
model-dependent uncertainties. Polarization propagators were derived in extended RPA approximation, including higher order many-body polarizations effects by particle and hole density- and energy-dependent in-medium self-energies. The approach is open for extension of various kind. An obvious one is the add more resonances, e.g. the spectrum of higher-lying $P_{11}$ and $P_{33}$ and the other P-wave states, $P_{13}$ and $P_{31}$. The roles of S-wave and higher spin-resonances are largely unexplored although they are considered in transport-theoretical models of CC reactions, e.g. \cite{Buss:2011mx}.

Peripheral heavy ion CC reactions at relativistic energies have a large, yet to be explored research potential for investigations of the transitional region of quasi-elastic to resonance excitations. Combined with pion spectroscopy as under preparation for the upcoming FAIR@GSI facility, data of a new quality will give insight into the spectral region of sub-nucleonic degrees of freedom.

\newpage
\appendix

\section{Spin and Isospin Structures of Propagators in Asymmetric Nuclear Matter}\label{app:SpinIso}
In particle-spin representation, the one particle-one hole (1ph) propagators are given as
\be\label{eq:G0mm}
\mathcal{G}^{(0)}(\omega,\mathbf{q})=\sum_{f_pf_h}|f_pf_h\ran G^{(0)}_{f_pf_h}(\omega,\mathbf{q})\lan f_pf_h|
\ee
where $|f_pf_h\ran$ are the spin-isospin wave functions of the particle-hole states. Hole states are always given by nucleons $N=p,n$ with spin and isospin $s_N,t_N=\frac{1}{2}$ where proton and neutron are carrying isospin charges $q_p=+1$ and $q_n=-1$, respectively. Particle states are either nucleons or excited states $N^*$ of the nucleon. The resonances $N^*$ considered here include $s_{N^*},t_{N^*}=\frac{1}{2}$ states like the Roper resonances, $P_{11}(1440)$ and $s_{N^*},t_{N^*}=\frac{3}{2}$ $P_{33}$-resonances, i.e. the Delta-states. For spin-scalar 1ph self-energies and spin-saturated matter, as assumed here, the propagators are independent of magnetic spin-quantum numbers. Under that proviso, response functions for spin transitions of different kind do not mix, i.e. transitions with spin-transfer $S=0$ decouple from those with $S=1$. Likewise, longitudinal and transversal spin modes decouple accordingly.

The solution of the Dyson-equation for the purely nucleonic excitations requires to evaluate polarization propagators of the kind
\be
\Pi^{(0)}_{a\alpha,b\beta}=\lan 0|\sigma_\beta \tau_b\mathcal{G}^{(0)}\tau_a\sigma_\alpha |0\ran .
\ee
where in spherical representation $\alpha,\beta=0,\pm 1$ and accordingly for the isospin operators.

We consider first the spin degrees of freedom only.
From Eq.\eqref{eq:G0mm} and by means of the results derived in App. \ref{app:2ndQ}, such expressions are readily evaluated. Under the given constraints, the summation over the magnetic spin quantum numbers can be performed, leading to $\delta_{\alpha\beta}$ and only the pieces diagonal in the spin-operators survive which, in fact, are independent of the projection quantum numbers. Since in our convention, the reduced matrix element for the unity operator $\mathbf{1}$ coincides in value with, $R^{(\mathbf{1})}_{\frac{1}{2}\frac{1}{2}}=R^{(\sigma)}_{\frac{1}{2}\frac{1}{2}}=\sqrt{2}$, we can write for configurations involving spin-$\frac{1}{2}$ particle and hole states
\be
\Pi^{(0)}_{N^*N}=R^{(\sigma/\mathbf{1})2}_{\frac{1}{2}\frac{1}{2}} G^{(0)}_{f_{N^*_p}f_{N_h}}=2G^{(0)}_{f_{N^*_p}f_{N_h}}
\ee
where now $N^*$ includes nucleon particle states as well.

The complexities of the composition of the system under consideration are still contained in the reduced propagators. Different from the widely cited and used expression found in \cite{FW:1971}, here we have to account for differences in particle and hole masses which change the momentum structure of the propagators. Moreover, in asymmetric nuclear the different content of protons and neutrons leads to a spontaneous breaking of isospin symmetry, reflected in two essential entities: Dynamically, isovector self-energies appear in the particle and the hole sector and statistically, protons and neutrons occupy Fermi-seas of different Fermi-momenta, $k_{F_p}\neq k_{F_n}$. These two kinds of effects are showing up especially in charge-exchange excitations which connect different baryon sectors. Either of those effects inhibits to perform the summation over isospin projections in closed form.

Including isospin, we find for configurations with isospin-$\frac{1}{2}$ particle and hole states
\bea
\Pi^{(0)}_{N^*N}&=&R^{(\sigma/\mathbf{1})2}_{\frac{1}{2}\frac{1}{2}}R^{(\tau/\mathbf{1})2}_{\frac{1}{2}\frac{1}{2}}
\left(t_p q_p t_h -q_h|\kappa q  \right)\left(t_p q_p t_h -q_h|\kappa' q'  \right)
P^{(N^*N)}_{k_{F_p}k_{F_h}}\\
&=&4\left(t_p q_p t_h -q_h|\kappa q  \right)\left(t_p q_p t_h -q_h|\kappa' q'  \right)
P^{(N^*N)}_{k_{F_p}k_{F_h}}
\eea
allowing isospin transitions with $\kappa=0,1$ where charge-exchange transitions proceed by $\kappa=1$ isovector transitions only.
The remaining propagator, stripped off spin and isospin matrix elements, is evaluated as retarded propagator
\be
P^{(N^*N)}_{k_{F_p}k_{F_h}}(\omega,\mathbf{q})=\int \frac{d^3k}{(2\pi)^3}
\left(X_{ph}(\omega,\mathbf{q},\mathbf{k})-Y^*_{ph}(\omega,\mathbf{q},\mathbf{k})  \right)
\ee
where for cold (T=0) nuclear matter
\be
X_{ph}(\omega,\mathbf{q},\mathbf{k})=\frac{\Theta(|\mathbf{k}+\mathbf{q}|^2-k^2_{F_p})\Theta(k^2_{F_h}-k^2)}
{E_p(k+q)-E_h(k)-\omega-i\Gamma_{ph}(\omega)/2}
\ee
and $Y^*_{ph}(\omega,\mathbf{q},\mathbf{k})=X_{ph}(-\omega,\mathbf{q})$. The energies are taken to include rest masses, mean-field and dispersive self-energies, e.g.:
\be
E_p(\mathbf{k}_p)=m^*_p+\frac{k^2_p}{2m^*_p}+U_p+Re(\Sigma_p(\omega))
\ee
and correspondingly for $E_h$. The imaginary parts of the dispersive self-energies are contained in the width
\be
\Gamma_{ph}(\omega)\approx -2Im(\Sigma_p(\omega)-\Sigma_h(\omega))
\ee
where we have neglected contributions due to the fact that we are dealing with particle-hole configurations.
Effective masses and self-energies are functions of the density and depends also on the proton and neutron content.

Defining
\bea
&&x(k,q,k_{F_p})=\frac{k^2+q^2-k^2_{F_p}}{2kq}\\
&&z(k,q,\omega)=\frac{1}{2kq}\\
&&\times \left(2m^*_p\left(-E_h(k)+m^*_p+U_p+Re(\Sigma_p))-\omega -\frac{i}{2}\Gamma_{ph}\right) +k^2+q^2\right)\nonumber
\eea
the integral over $t=\cos{\theta_{kq}}$  can be done analytically, leading to the Lindhard-functions
\bea
\phi_{k_{F_p}k_{F_h}}(\omega,q,k)&=&\frac{1}{2}\int_{-1}^{+1}dt\frac{\Theta(x+t)}{z+t} \\
&=&
\frac{1}{2}\left(\log{\left(\frac{z+1}{z-x}\right)}\Theta(x+1)-\log{\left(\frac{z-1}{z-x}\right)}\Theta(x-1)   \right).\nonumber
\eea
and
\be
P^{(N^*N)}_{k_{F_p}k_{F_h}}(\omega,\mathbf{q})=
-\frac{m^*_p}{2q\pi^2}\int_{0}^{k_{F_h}}dk k\left(\phi_{k_{F_p}k_{F_h}}(+\omega,\mathbf{q})-\phi_{k_{F_p}k_{F_h}}(-\omega,\mathbf{q})  \right).
\ee
Even the remaining integral can be done in closed form but we refrain from displaying the quite lengthy expression.

The particle Fermi-momentum is non-zero only for purely nucleonic excitations. While for $nn^{-1}$ and $pp^{-1}$ modes we have $k_{F_p}=k_{F_h}$ and $m^*_p=m^*_h$ a completely different situation is encountered in the $np^{-1}$ and $pn^{-1}$ charge exchange channels. In asymmetric nuclear matter, the particle and hole Fermi-momenta and effective masses will be different which is also true for the self-energies.

The same strategy can also be applied to the channels involving Delta-particle states:
\bea
\Pi^{(0)}_{DN}&=&R^{(S)2}_{\frac{3}{2}\frac{1}{2}}R^{(T)2}_{\frac{3}{2}\frac{1}{2}}
\left(t_D q_D t_h -q_h|1 q  \right)\left(t_D q_D t_h -q_h|1 q'  \right)
P^{(DN)}_{k_{F_h}}\\
&=&\frac{16}{9}\left(t_D q_D t_h -q_h|1 q  \right)\left(t_D q_D t_h -q_h|1 q'  \right)
P^{(DN)}_{k_{F_h}}
\eea
where
\be
P^{(DN)}_{k_{F_h}}(\omega,\mathbf{q})=
-\frac{m^*_D}{2q\pi^2}\int_{0}^{k_{F_h}}dk k\left(\phi_{DN}(+\omega,\mathbf{q})-\phi_{DN}(-\omega,q)  \right).
\ee

\section{Spin and Isospin Formalism in Second Quantization}\label{app:2ndQ}
In second quantization, the spin-operators are
\bea
\sigma_\mu&=&\sum_{m_1m_2}\lan m_1|\sigma_\mu|m_2\ran a^+_{m_1}a_{m_2}\label{eq:sigma1}\\
          &=&\sum_{m_1m_2}R^{(\sigma)}_{ss}(-)^{s_2+m_2}(s_1m_1s_2m_2|1\mu)a^+_{m_1}a_{-m_2}\label{eq:sigma2}
\eea
where $s_1=s_2=s=\frac{1}{2}$ and the Wigner-Eckart theorem was used. The reduced matrix element of the operator $\bm{\sigma}$ of tensorial rank $\lambda=1$ is defined as
\be \label{eq:Rss}
R^{(\sigma)}_{ss}=\frac{1}{\hat{\lambda}}\lan s||\bm{\sigma}||s\ran=\frac{2}{\sqrt{3}}\sqrt{s(s+1)(2s+1)}_{|s=\frac{1}{2}}=\sqrt{2}.
\ee
Eq.\eqref{eq:sigma2} shows that the components $\sigma_\mu$ mediate spin-dipole excitations $(\lambda,\mu)=(1,\mu$. In the same manner, we find for the isospin rank-1 operators
\bea
\tau_\mu&=&\sum_{q_1q_2}\lan q_1|\sigma_\mu|q_2\ran c^+_{q_1}c_{q_2}\label{eq:tau1}\\
          &=&\sum_{q_1q_2}R^{(\tau)}_{tt}(-)^{t_2+q_2}(t_1q_1t_2q_2|1\mu)c^+_{q_1}c_{-q_2}\label{eq:tau2}
\eea
where the reduced isospin matrix element $R^{(\tau)}_{tt}$ is defined in analogy to Eq.\eqref{eq:Rss}.

Excitations of Delta particle-nucleon hole configurations involve transitions between spin-/isospin-$\frac{1}{2}$ and spin-/isospin-$\frac{3}{2}$ baryons for the so-called transition spin/isospin formalism with rank-1 operators $\mathbf{S}$ and $\mathbf{T}$, respectively,  is used, see e.g. \cite{Brown:1975di}. The transition spin operator $\mathbf{S}$ - and correspondingly $\mathbf{T}$ - is defined by the matrix elements of nucleon ($s_N=\frac{1}{2}$) to Delta ($s_D=\frac{3}{2}$) transitions
\be
\lan s_D m_D|S_r|s_Nm_N\ran \equiv \left(s_N m_N 1 r|s_Dm_D \right)
\ee
given on the right hand side by a Clebsch-Gordan coefficient.
By comparison to the standard form of the Wigner-Eckart theorem, we obtain the reduced matrix element
\be
\lan \frac{3}{2}||\mathbf{S}||\frac{1}{2}\ran=\sqrt{2s_D+1}_{|s_D=\frac{3}{2}}=2
\ee
and
\be\label{eq:RSDN}
R^{(S)}_{s_Ds_N}=\frac{\lan \frac{3}{2}||\mathbf{S}||\frac{1}{2}\ran}{\sqrt{2\lambda+1}}_{|\lambda=1} =\frac{2}{\sqrt{3}}
\ee
As before, we may use the formalism of second quantization to obtain
\bea
S_r&=&\sum_{m_Dm_N}\lan s_D m_D|S_r|s_Nm_N\ran a^+_{m_D}a_{m_N} + h.c. \label{eq:Sr1}\\
&=&\sum_{m_Dm_N}R^{(S)}_{s_Ds_N}\left(\left( s_D m_Ds_Nm_N|1 r\right) (-)^{s_N+m_N} a^+_{m_D}a_{-m_N} + h.c.\right).
\label{eq:Sr2}
\eea
Accordingly, we proceed for the isospin transition operator $\mathbf{T}$. The reduced matrix element for $t_N=\frac{1}{2}$ and $t_D=\frac{3}{2}$, respectively, is
\be\label{eq:RTDN}
R^{(T)}_{t_Dt_N}=\frac{\sqrt{2t_D+1}}{\sqrt{2\lambda+1}}_{|t_D=\frac{3}{2},\lambda=1}=\frac{2}{\sqrt{3}}
\ee
and
\bea
T_r&=&\sum_{q_Dq_N}\lan q_D q_D|T_r|t_Nq_N\ran a^+_{q_D}a_{q_N} + h.c. \label{eq:Tr1}\\
&=&\sum_{q_Dq_N}R^{(T)}_{t_Dt_N}\left(\left( t_D q_Dt_Nq_N|1r\right) (-)^{t_N+q_N} c^+_{q_D}c_{-q_N} + h.c.\right). \label{eq:tr2}
\eea


\end{document}